# THE LIGHT CAME IN 1905[*]

**Jozef T. Devreese**

*Theoretische Fysica van de Vaste Stoffen, Universiteit Antwerpen,
U.342 / Campus Groenenborger, Groenenborgerlan 17, Antwerpen-2020, België
and
eiTT/COBRA Inter-Universitair Onderzoeksinstituut,
Technische Universiteit Eindhoven,
NL-5600 MB Eindhoven, Nederland*

Our present Worldview can not be imagined without the seminal ideas of Albert Einstein.

## Introduction

Today, celebrating the centennial of Albert Einstein's 1905 papers, we realize that our present worldview is unimaginable without his work. In this *World Year of Physics*, the European Physical Society, UNESCO and UN have decided to celebrate physical sciences on the world scale. With the aim, "of developing trans-border and European relations", the French Physical Society and the Belgische Natuurkundige Vereniging / Société Belge de Physique decided to merge their congresses, and to organize the present Joint SFP-BPS Scientific Meeting, in Lille /Rijsel.

Einstein's seminal works on light quanta, relativity, fluctuation-dissipation and Brownian motion have determined many of the central themes of physics up to the present. The subjects shown in Fig. 1, have been selected from the talks at the 13$^{th}$ General Conference of EPS held in Bern in July of this year and illustrate the lasting impact of Einstein's works.

In the present talk, I will restrict myself to selected topics and aspects of Einstein's work and its significance. The time limitations force me to follow Einstein's advice: *"Simplify as much as possible… but not more."*

The following subjects will be discussed: Einstein's Compass, Annus Mirabilis, General Relativity, Bose-Einstein Statistics and BEC, and Beyond Einstein.

---

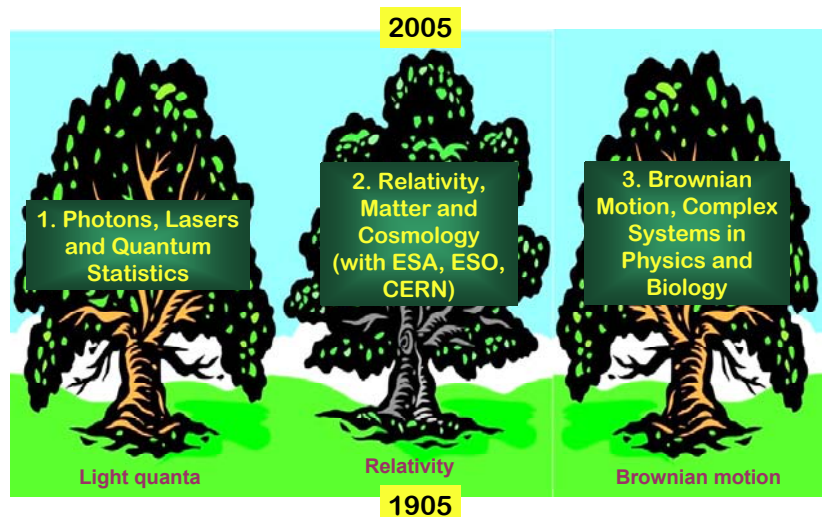

Fig. 1. Topics of the program of the 13<sup>th</sup> General Meeting of the European Physical Society (July 11-14, 2005, Bern, Switzerland).

# 1. Einstein's compass

It is striking how Einstein, in his work, seems to have been guided by general principles. A kind of "compass" seems to be directing him towards his achievements. From his biography [1] we know about "the two miracles" that have strongly influenced young Albert:
- the encounter with a compass and
- the encounter with a geometry book.

He remembered in his *Autobiographisches*: "*I encountered a wonder ... as a child of 4 or 5 years when my father showed me a compass. That this needle behaved in such a determined way did not fit into the way of incidents at all... There must have been something behind things that was deeply hidden*". So, this was the first "miracle".

About his "*holy geometry book*", the second "*miracle*", he wrote: "*At the age of 12 I experienced a second wonder of a very different kind: a booklet dealing with Euclidian plane geometry that came into my hands at the beginning of a school year... The clarity and certainty of its contents made an indescribable impression on me*". Such private experiences contributed far more to Einstein's growth than did formal schooling.

One of the most popular myths about Einstein is that he failed at school. This – however – is quite wrong. In Einstein's certificate received from the Canton school in Aarau, the lessons, he was less interested in, can easily be detected (see Fig. 2). But the average grade in his certificate was a 5 out of maximal six, i.e. the grade "good". He also obtained 6/6 for Algebra, Geometrie, Darstellende Geometrie, Physik.

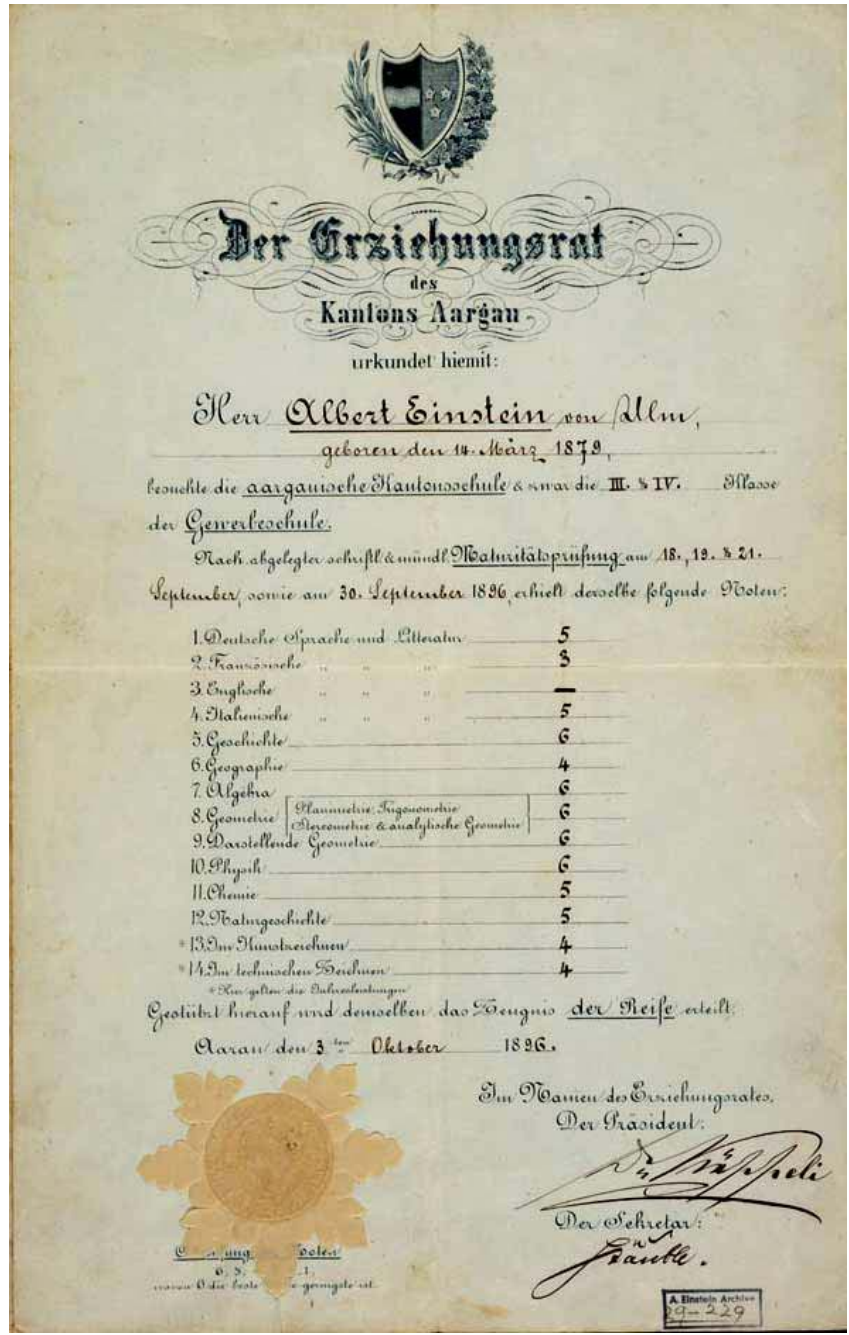

Fig. 2. Einstein's certificate received from the Canton school in Aarau.
(From: http://www.wilsonsalmanac.com/images2/einstein_diploma.jpg)

From 1902 till 1909 Einstein worked at the patent office in Bern, as a "technical expert; third class" as is well known. Regarding his activities there, he wrote: "*A practical profession is a salvation for a man of my type; an academic career compels a young man to scientific production, and only strong characters can resist the temptation of superficial analysis*". This statement of Einstein is significant to us, in a time where counting the *number* of publications and the *number* of citations has become the *sacred vogue*.

Remembering his time in the Bern patent office, Einstein wrote: *"I was sitting in a chair in the patent office at Bern when all of sudden a thought occurred to me: 'If a person falls freely he will not feel his own weight'. I was startled. This simple thought made a deep impression on me. It impelled me toward a theory of gravitation"*. Later, he called this *"the happiest thought of my life"*.

When you consider that Einstein was not connected with any university, or physics laboratory or physicist when he conceived and published his *"five papers that shook the world"*, it really makes you think. He wrote "Nobel-prize papers" as a PhD student, without a thesis advisor.

***"Do not follow trends; set trends"***

Writing to Willem de Sitter in 1916, Einstein stated: "*I am driven by my need to generalize* [*mein Verallgemeinerungsbedürfnis*]." That *Verallgemeinerungsbedürfnis* was clearly a driving force, a *compass,* orienting Einstein's trajectory. He worked to generalize the special theory of relativity to include gravitational fields and to produce what he first called, in an article of 1913, the *generalized* relativity theory. And, finally, Einstein threw himself into the search for a *unified field theory*, the ultimate generalization.

John Ellis [2] says the following about Einstein's quest for unification: "*It seems to me that the real significance of Einstein's quest for unification lies in its quixotic ambition. Einstein, more than any of his contemporaries, put unification on the theoretical map and established it as a respectable intellectual objective. Even if we do not have all the necessary theoretical tools or experimental information, unification is the "holy grail" towards which our efforts should be directed.*"

Einstein tackled a wide spectrum of subjects ranging from the specific heat of solids to the meandering of rivers, including the famous (so called) "paradox" of Einstein, Podolsky and Rosen (EPR). In 1935, they proposed [3] a thought experiment to demonstrate that quantum mechanics is incomplete. The EPR "paradox" stimulated attempts to verify the foundations of quantum mechanics experimentally, for example the photon-polarization measurements by Aspect, Dalibard, and Roger in 1982 [4]. The problems brought up by EPR not only reflect academic interest, but they hold the promise of practical applications for quantum teleportation, quantum computing and quantum cryptography.

Einstein had a keen interest in philosophy and in the arts (he played the violin, as is well known). He had strong opinions on pacifism, supranationalism and civil liberties. The media turned him into a public figure and he interacted with celebrities like Gandhi, Tagore, Nehru, Freud, Chaplin, Ben Gurion, Churchill… to name a few. His conversations with Tagore (1919 and 1930) were recorded and are centered on causality, "*objective reality*" and Western versus Eastern music. From those conversations, I select the following statement by Einstein, about objective reality: *"I believe, for instance, that the Pythagorean theorem in geometry states something that is approximately true, independent of the existence of man"*.

## 2. Annus Mirabilis

Next we come to 1905, the by now well publicized *Annus Mirabilis*, when Einstein blazed new paths for science and mankind. Wilczek's appreciation of the 1905 papers is: *"Einstein's work on Brownian motion would have merited a sound Nobel prize, the photoelectric effect a strong Nobel prize, but special relativity and $E = mc^2$ were worth a super-strong Nobel prize"* (cited from [5]).

The insights which Einstein derived – with his marvelous intuition – were, in a certain sense, prepared by interplay of earlier discoveries. Einstein refers e.g. to two of his great predecessors, as follows: *"There is no doubt, that the special theory of relativity, if we regard its development in retrospect, was ripe for discovery in 1905. Lorentz has already observed that for the analysis of Maxwell's equations the transformations which later were known by his name are essential, and Poincaré had even penetrated deeper into these connections."*

### *The miracles of 1905*

**March 1905.** Einstein sent to *Annalen der Physik,* a paper with a new understanding of the structure of light. He argued that light can act as though it consists of quanta.

**April 1905.** Einstein completed his PhD thesis, *"Eine neue Bestimmung der Moleküldimensionen"* ("On a new determination of molecular dimensions"), published in 1906 in *Annalen der Physik.*

**May 1905 & December 1905.** *Annalen der Physik* received further work from Einstein on the kinetic theory, with his explanation of Brownian motion.

**June 1905.** Einstein sent to *Annalen der Physik* a paper on electromagnetism and motion, which represents the special relativity theory.

**September 1905.** Einstein reported a remarkable consequence of his special theory of relativity: if a body emits a certain amount of energy, then the mass of that body must decrease by a corresponding amount.

Table 1 displays a citation count for Einstein's works of 1905. Of course, the significance of the given figures should not be overestimated: they take into account only the *explicit references to the papers* and only since 1972. Obviously, the number of publications, where Einstein's results are actually exploited without a reference to his papers or even to his name, is by orders of magnitude larger. Nevertheless, the table may give an idea of how often Einstein's works of 1905 are cited relative to each other.

Among these works, the thesis appears to be the most cited. This is because the thesis, dealing with rheological properties of particle suspensions, contains results, which have an extraordinary range of applications relevant to the construction industry, polymer technology, dairy industry and ecology… Einstein tops the table of most-cited pre-1930 papers, according to Werner Marx and Manuel Cardona [6].

Table 1. Citation count for Einstein's works of 1905.

**Einstein – 1905**

| | | |
|---|---|---|
| Annalen der Physik 17, 132 (1905) | Photoelectric Effect | 396 |
| Annalen der Physik 17, 549 (1905) | Brownian Motion | 1552 |
| Annalen der Physik 17, 891 (1905) | Special Relativity | 737 |
| Annalen der Physik 18, 639 (1905) | $E = mc^2$ | 108 |
| Annalen der Physik 19, 289 (1906) (Thesis Work) | Molecular Size | 1643 |

**Explicit-citation count**
(According to ISCI, 1972 – August 20, 2005)

*Light quanta. The photoelectric effect*

Einstein's 1905 paper on light quanta endowed Max Planck's quantum hypothesis with physical reality. *"According to the assumption considered here, when a light ray starting from a point is propagated, the energy is not continuously distributed over an ever increasing volume, but it consists of a finite number of energy quanta, localized in space, which move without being divided and which can be absorbed or emitted only as a whole."*

Arguably this constitutes his most revolutionary discovery. His idea – that light consists of indivisible particles (quanta) – which challenged the electromagnetic theory of light, was initially rejected by contemporary physicists (Millikan, Bohr and – very explicitly – Planck). Even in 1922, when Einstein received the 1921 Nobel prize "especially for his discovery of the *law of the photoelectric effect*", Niels Bohr, recipient of the 1922 prize, said in his Nobel address: *"The hypothesis of light-quanta is not able to throw light on the nature of radiation"*.

Only after Compton's 1923 X-ray scattering experiment did physicists finally accept Einstein's idea. In 1926 Einstein's quantum of light was named "photon" by Gilbert Lewis, a chemist.

Bose, in 1920, derived Planck's formula starting from "bare essentials" and making use of the corpuscular nature of light. He invoked statistical rules using the indistinguishable nature of Einstein's light quanta and implied the nonconservation of photons. Bose intuitively realized that a factor "2" in this formula, is connected with the polarization of the light quantum. He wrote of this polarization -factor "2": *"it seems required"*. Only later a further foundation for this factor "2" in terms of the "photon spin" could be given in QED. In modern field theory the photon is described as the exchange particle of the electromagnetic interaction and it was the first example of a gauge boson, eventually leading to the gauge theories that form today's standard model. One of such theories will be presented at the Joint SFP-BPS Scientific Meeting by Prof. Englert.

The impact today of Einstein's paper on the photoelectric effect is illustrated by Fig. 3.

## Photoelectric Effect Today

- **Solar panels**
- **Photomultipliers**
- **Photoelectric cells**
- **Night vision devices**
- **Image converters**
- **Image storage tubes**
- **Photocathodes**
- **Photochemistry**
- **Photoemission electron microscopy**
- **Photoemission spectroscopy**
- **Single-photon sources**
- **Cavity quantum electrodynamics**
- **Nonlinear optics**
- **…**

Technological applications: Night vision devices...

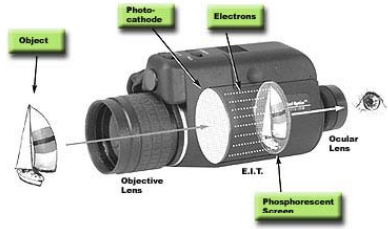

Scientific applications: Probe of electron structure of solids: semiconductors, high-$T_c$ superconductors, polymers…

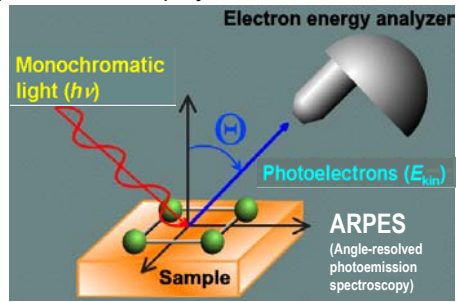

http://www.ifw-dresden.de/iff/11/arpes/arpes.htm

Fig. 3. Impact of Einstein's paper on the photoelectric effect.

*Brownian motion*

Einstein's paper of May 1905 demonstrated how Brownian motion can prove the existence of molecules and atoms. In his PhD thesis Einstein also discovered the first example of a fluctuation-dissipation theorem, which is also exploited in the paper on Brownian motion. In his later years, immersed in the search for a "Unified Field Theory" based on his general theory of relativity, Einstein himself dismissed his work on Brownian motion as unimportant. One century later, as can be seen from Fig. 4, Einstein's ideas about Brownian motion flourish in modern science. They have impact from physics to biology and to the latest wonders of nanotechnology... And *econophysics* is applied daily in Wall Street. Concepts such as random walk process, Markov process, Brownian motion, underlie modern economic models.

## Brownian Motion Today

- Physics of noise
- Seismology
- Diffusion, including diffusion on crystalline surfaces
- Polymers in aqueous solutions
- Econophysics
- Relativistic Brownian motion*
- Life sciences, including biological physics
- Industrial chemistry & chemical engineering
- Physical properties of oils
- Particle image velocimetry
- Electrophoresis
- Hadron-production processes
- Compressible fluids
- Molecular motors
- Directional crystallization
- Granular matter
- Scanning microscopy
- …

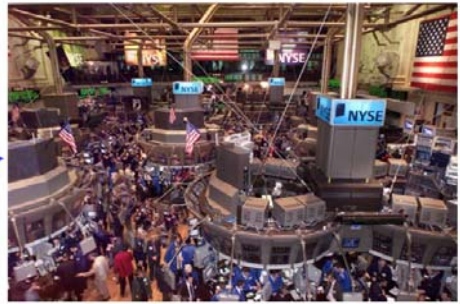

Not a physics hall, but the New York Stock Exchange

Applications in technology, marketing, … and science

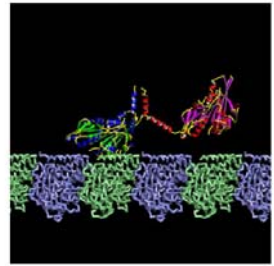

Model of kinesin dimer walking along a microtubule protofilament
http://www.mpasmb-hamburg.mpg.de/ktdock/index.html#kinwalk

Fig. 4. Impact of Einstein's paper on Brownian motion.

### *Special relativity*

Although, in Einstein's own words, *"the special theory of relativity... was ripe for discovery in 1905"*, apparently it needed an unbiased newcomer to take the final step.

In his Kyoto address, in 1922, Einstein said on the theory of special relativity: *"My solution was really for the very concept of time, that is, that time is not absolutely defined but there is an inseparable connection between time and the signal velocity... Five weeks after my recognition of this, the present theory of special relativity was completed."*

Special relativity combined with quantum mechanics gave rise to the Dirac equation, to quantum field theory, to the spin-statistics connection and to the discovery of anti-matter.

So far, Lorentz invariance has survived all experimental tests…

In his 1905 paper, Einstein discovered that *"The mass of a body is a measure of its energy-content"*. Today, $E = mc^2$, one of the best-known equations, is also a catch-phrase probably as familiar to the public as any line from Shakespeare or Voltaire. The symbol $c$ (for the velocity of light in vacuum; from "celeritas") was introduced by Einstein in 1907. Before that date he used the symbol $V$, which had been introduced by Maxwell.

That the speed of light in vacuum is the same in all inertial frames, remains surprising, even today. Einstein's theory of relativity and the atomistic model –with their counter-intuitive

concepts- reverberate in general culture and even with the general public. Some works of art, like e.g. Dali's *Raphael Madonna at maximum speed* seem to reflect this fact. (Fig. 5).

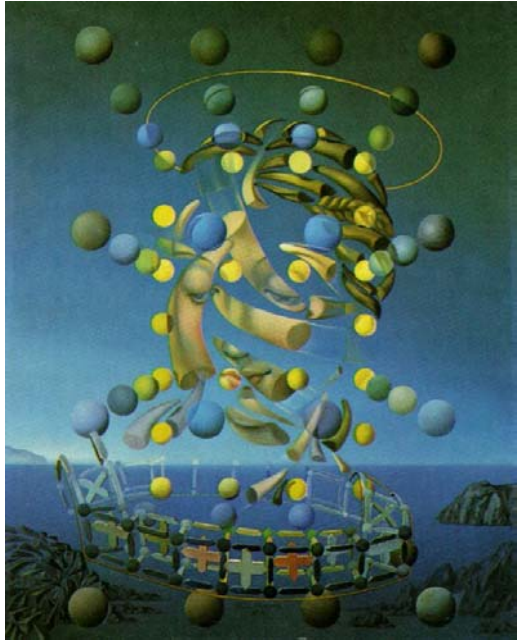

Fig. 5. Salvador Dali, *Maxima Velocidad de la Madona de Rafael* (1954)
(From http://www.dali-gallery.com/html/paintlst.htm.)

## 3. General relativity

It took Einstein 8 years, from 1907 to 1915, to complete the general theory of relativity and to arrive at his field equations. In that effort, he was strongly guided by his *equivalence principle*. Success in Einstein's theoretical work on gravity was sealed – after an immense intellectual effort – in November 1915. On December 10 (1915) just after this achievement, Einstein wrote to his friend Besso that he was "*zufrieden aber ziemlich kaputt*".

Today "General relativity is at the heart of cosmological models", which originate from the works of Einstein, de Sitter, Friedman, Lemaitre… Einstein's new equations of gravitation possess a great logical simplicity, despite their mathematical form, which was quite unfamiliar at the time.

As Wilczek formulates it [7], *"The central equation of general relativity theory,*

$$R_{\mu\nu} - \frac{1}{2} g_{\mu\nu} R = T_{\mu\nu}$$

*(in appropriate units), equates the curvature of spacetime [lhs] to the energy–momentum of matter [the content of spacetime, rhs]. Einstein referred to the left-hand side as a palace of gold, and to the right-hand side as a hovel of wood, thus expressing his ambition to make improvements on the right-hand side, to root it in concepts of depth and beauty comparable to*

*Riemannian geometry."* Here, $R_{\mu\nu}$ is the Ricci tensor, $R$ is the Ricci scalar, $g_{\mu\nu}$ is the metric tensor and $T_{\mu\nu}$ is the energy-momentum tensor.

To what degree did general relativity find experimental confirmation?

In November 1919 Eddington announced to the world that observations of a solar eclipse in May 1919 supported Albert Einstein's general theory of relativity (see Table 2). That announcement was one of the most influential events of 20th-century science, and was largely the cause of Einstein's fame.

Table 2. Deviations of the stars, as observed during a solar eclipse in May 1919, in comparison with the predictions of the general relativity.
(From http://www.ivorix.com/en/einstein/contents/ap03.html.)

**The observed and calculated deviations of the stars (in seconds of arc)**

| Number of the Star. | First Co-ordinate. | | Second Co-ordinate | |
|---|---|---|---|---|
| | Observed. | Calculated. | Observed. | Calculated. |
| 11 | −0·19 | −0·22 | +0·16 | +0·02 |
| 5 | +0·29 | +0·31 | −0·46 | −0·43 |
| 4 | +0·11 | +0·10 | +0·83 | +0·74 |
| 3 | +0·20 | +0·12 | +1·00 | +0·87 |
| 6 | +0·10 | +0·04 | +0·57 | +0·40 |
| 10 | −0·08 | +0·09 | +0·35 | +0·32 |
| 2 | +0·95 | +0·85 | −0·27 | −0·09 |

Later some claimed that the accuracy of the observations could be insufficient for a quantitative support of the theory. Further progress in the field is illustrated by the plot of the coefficient $(1+\gamma)/2$ (see Fig. 6). The parameter $\gamma$ measures how much spacetime-curvature is produced by unit rest mass. Its value differs for different metric theories. Its general-relativity value is unity. You see here how improved observations have led to .02 % agreement with general relativity.

However, we are reminded of Chandrasekhar's caveat [8]: *"But all of (the above) effects relate to departures from the predictions of the Newtonian theory by a few parts in a million, …And, so far, no predictions of general relativity, in the limit of strong gravitational fields, have received any confirmation, and none seem likely in the foreseeable future."*

You certainly have heard about LIGO (Laser interferometer gravitational wave observatory)] and about LISA (Laser Interferometer Space Antenna). The objective of those projects is precisely to probe the strong gravity limit, contemplated by Chandrasekhar.. Several general-relativity experiments deal with testing the Equivalence Principle (i.e. $m_{gravity} = m_{inertial}$; the "weak equivalence principle"). The evolution of the experimental *accuracy* for the upper bound of the relative difference between the gravitational and the inertial masses is shown in Fig. 7.

> A remark about the history of physics is in order here: often Galileo is represented as the first to have tested the equivalence principle but he was preceded e.g. by Benedetti and by Simon Stevin. There is also no evidence that Galileo ever dropped two balls of different weights off the leaning tower of Pisa. That *Physics World* proclaimed this fictitious PISA-tower experiment as the most beautiful physics experiment in history, is an outstanding example of *myth* proving stronger than *fact,* even – sometimes – for physicists!

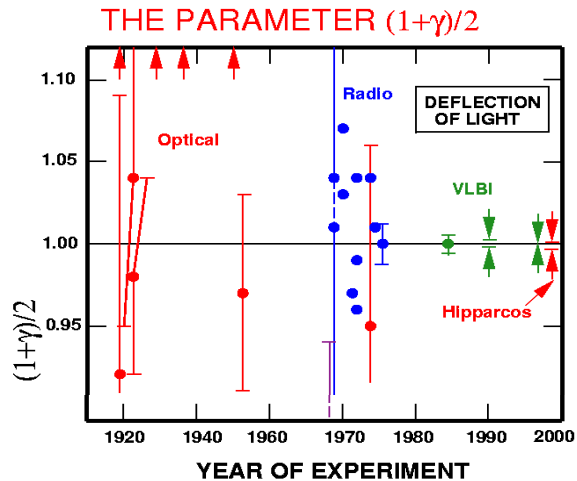

Fig. 6. Plot of the coefficient (1+γ)/2. The arrows denote anomalously large values from early eclipse expeditions. The development of very-long-baseline radio interferometry (VLBI) produced greatly improved determinations of the deflection of light (1969–1975). A recent series of VLBI quasar and radio galaxy observations yield an agreement with GR to 0.02 percent. *Hipparcos* denotes the ESA astrometric satellite, which has reached 0.1 percent. (After: http://arxiv.org/PS_cache/gr-qc/pdf/9811/9811036.pdf.)

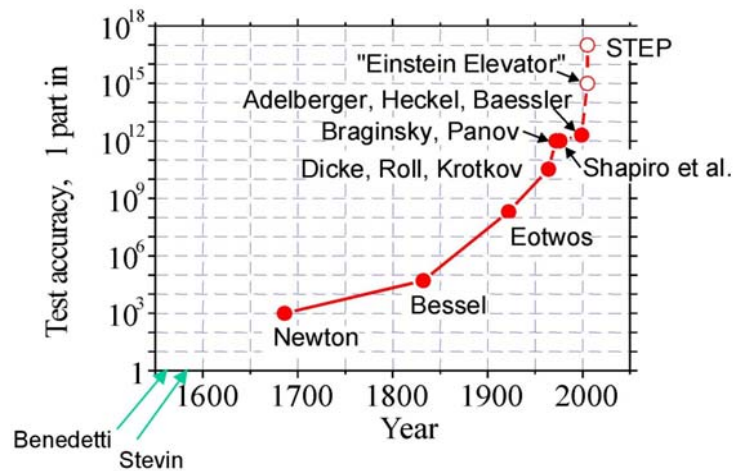

Fig. 7. Progress in testing the equivalence of inertial and gravitational mass.
(After: http://cfa-www.harvard.edu/equiv/,
http://cosmology.berkeley.edu/~miguel/GravityEtCetera/GravityPages/STEPTheory.html .)

In 1993, the Nobel Prize in Physics was awarded to Hulse and Taylor for their 1974 discovery of a double neutron star system containing a pulsar [PSR1913+16]. The two neutron stars orbit around a common centre of mass. General relativity predicts that such a binary system will lose energy with time as orbital energy is converted into the radiation of gravitational waves. The observed decrease of the period of the pulsar orbital motion is in excellent agreement with that predicted by general relativity (see Fig. 8).

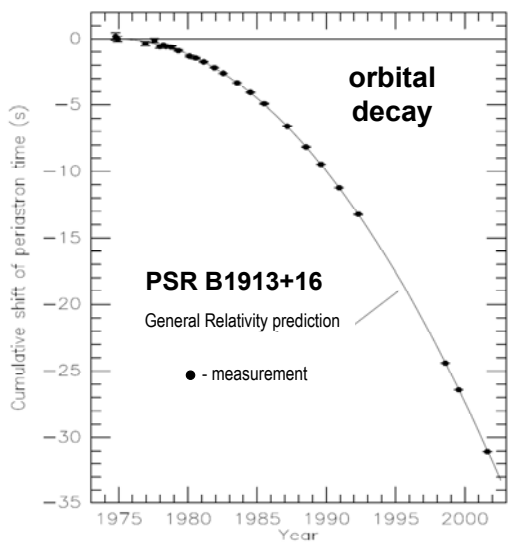

Fig. 8. Comparison of the observed decrease in the period of the pulsar orbital motion with general relativity predictions. (After [9].)

The observations of the "orbital decay" of this binary pulsar system provide the strongest (still indirect) evidence, to date, for the existence of gravitational waves. More recent experiments concern a pair of pulsars circling each other. This double pulsar system (PSR J0737-3039A and B) was discovered in 2003 by an international team of astronomers [10].

Although – at first sight - special and general relativity may seem to lack usefulness, they do have a measurable impact on everyday life. Special relativity predicts that clocks, moving at GPS-satellite orbital speeds with respect to a earth-centered non-rotating frame, will tick slower by about 7 microseconds/day than stationary ground clocks. General relativity predicts that the atomic clocks at GPS orbital altitudes will tick faster by about 45 microseconds/day than atomic clocks on the earth surface. The combination of these relativistic effects means that the on-board clocks would tick faster than clocks on the ground by about 38 microseconds per day. If these effects were not taken into account, errors in global positions would accumulate at the – maybe surprising – rate of about 10 kilometers per day.

## 4. Bose-Einstein statistics and BEC

Now I turn briefly to Bose-Einstein condensation. BEC was theoretically predicted by Einstein, who was inspired by Bose's paper on the statistics of photons. First observed – as you know- in 1995, BEC is one of the recent dramatic experimental confirmations of yet another fundamental physical phenomenon conceived by Einstein.

As communicated by *Associated Press* on August 20, 2005, the original manuscript of the Einstein paper "*Quantum theory of the monatomic ideal gas*" has been recently found in the archives of the *Lorentz Instituut voor Theoretische Fysica* of the *Universiteit Leiden*. A student working on his master's thesis, uncovered the manuscript among papers of Paul Ehrenfest.

Page 2 of the manuscript reports the prediction of the new state of matter now called the Bose-Einstein condensate (see Fig. 9). Considered as of Einstein's last great breakthroughs, it was published in the proceedings of the Prussian Academy of Sciences in Berlin in January 1925.

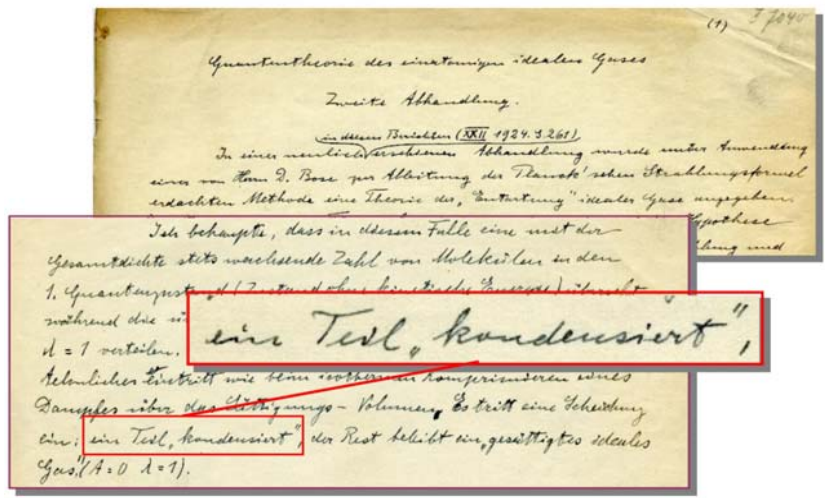

Fig. 9. Fragments of the original manuscript of Einstein's paper "Quantum theory of the monatomic ideal gas".
(From: http://www.lorentz.leidenuniv.nl/history/Einstein_archive/Einstein_1925_manuscript/.)

The phenomenon of Bose-Einstein condensation could not be observed until techniques were developed which allow us to reduce temperatures to about ~100 nK. Hänsch and Schawlow recognized in 1975 that laser light could be used to cool free atoms. The first successful experiments on laser cooling were performed around 1980 by Chu, Cohen-Tannoudji, nicknamed "*jongleur d' atomes*", and Phillips. They were awarded the Nobel Prize in Physics 1997 "*for the development of methods to cool and trap atoms with laser light*".

Isaac Silvera had pioneered the search for the observation of BEC with his experiments in the eighties (Amsterdam) on spin-polarized hydrogen. Cornell / Wieman and Ketterle, using very inventive and advanced methods (including laser cooling), finally managed to observe the Bose-Einstein condensation in 1995. BEC was achieved in alkali atom gases (originally with $^{87}$Rb atoms), in which the phenomenon can be studied in a very pure manner.

To allow for the unambiguous identification of *vortices* in BEC's, J. Tempere and the author [11] studied "colliding" BEC's and showed that *"edge dislocations"* appear in the interference patterns caused by these interpenetrating BEC's, if at least one BEC contains a vortex. In Fig. 10 you can see such an interference pattern, that can be obtained by sending light through the interpenetrating BEC's. We found that the fork-like structure in the interference pattern, which you see here, is the *signature*, the *fingerprint* of a *vortex* in a BEC. Our theory of 1998 was nicely confirmed by experiments [12,13] of both the Ketterle and the Dalibard teams in 2001 (see Fig. 11).

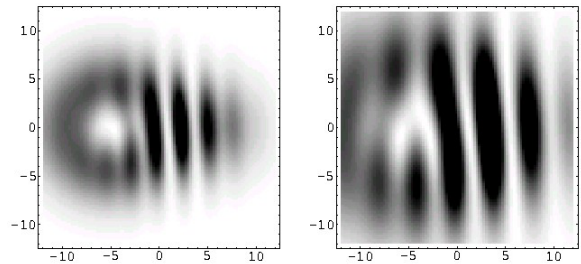

Fig. 10. Theoretical prediction for the time evolution of the density of interpenetrating, expanding condensates in the *x*, *y*-plane. The left condensate contains a vortex, the right condensate has none. The rhs panel corresponds to a later time as compared to the lhs panel. (From [11].)

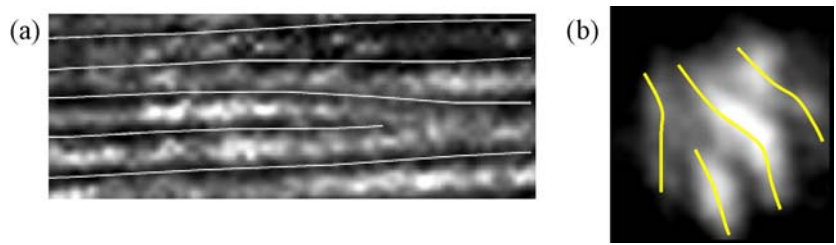

Fig. 11. Experimentally observed vortices in interfering BEC's. (Panel *a*: from [12], Panel *b*: from [13].)

An example of a fascinating experiment is taken from the work of Jean Dalibard and his associates. An important aspect of this work, which provides a clear observation of quantized vortices in BEC's, is that it proves that BEC is observed as a pure quantum phenomenon. In Fig. 12, you see 1 / 2 / 3 …vortices in a BEC. Einstein stressed that the condensation occurs without interactions.

Historically the concept of BEC played a role in the theoretical characterization of *phase transitions*. A very recent development concerns the observation of vortices (and evidence for superfluidity) in an ultracold Fermi gas of atoms [14].

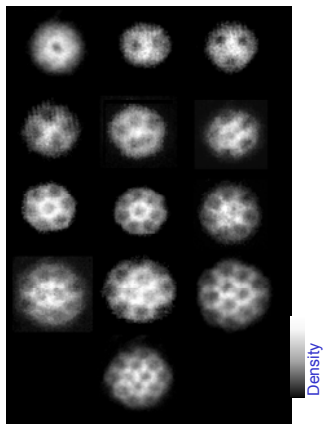

Fig. 12. Quantized vortices in BEC's.
(After: http://www.lkb.ens.fr/recherche/atfroids/anglais/activite_an.html)

## 5. Beyond Einstein

Many studies probe "beyond Einstein". *Quantum gravity* is a key area *beyond Einstein*, and *path integrals* turn out to be crucial here. Feynman's path integral formalism of quantum mechanics was originally invented for *non-relativistic* quantum mechanics. Gradually path integrals were generalized to be used in quantum field theory and later in *general relativity*.

Feynman has shown that *the description of a boson field with spin two* (cfr. the graviton field) *coupled to the energy-momentum field uniquely leads to Einstein's general relativity* (i.e. without geometry…).

The classical action (with Lagrangian $L$) is the key ingredient of the path integral. Rewriting the action in terms of the Lagrangian density $\mathcal{L}$ makes covariance transparent:

$$\int L\, dt \Rightarrow \int\int\int\int \mathcal{L}\, dx dy dz dt$$

Hawking contributed profound work using path integrals to *combine* quantum theory and general relativity and applied it to cosmology and to the study of black holes (see, e.g., [15,16]). In particular, he applied the path-integral method to the quantum mechanical study of a scalar particle moving in the background geometry of a Schwarzschild black hole. The amplitude for the black hole to emit a particle is expressed as a sum over paths [15]:

$$K(x,x') \sim \sum_{\text{paths}} e^{iS(x,x')/\hbar}$$

His *non-perturbative* approach leads to a relation between the probabilities for a black hole to emit a particle with energy $E$ and that to absorb a particle with the same energy:

$$W_{\text{Emission}} = e^{-2\pi E/\kappa} W_{\text{Absorption}}$$

$\kappa$ is the surface gravity of a black hole. Hawking's formula shows that a black hole will emit particles with a thermal spectrum characterized by a temperature, which is proportional to the surface gravity of the black hole. His work suggests a link between general relativity, quantum theory and thermodynamics… Our colleagues at CERN hope to create mini black holes with the L.H.C. – Large Hadron Collider.

## Concluding remark

Einstein had a great impact with the general public, he became an *icon* and was virtually *"canonised"*. Ever since "Einstein - 1905" it became clear that the world (even the measurable world and the laws describing it) defies common sense…

It is indeed a true measure of Einstein's legacy that: "*His name alone* symbolizes science" [17].

**Acknowledgments.** I would like to acknowledge the pleasant interactions I had with colleagues and collaborators during the preparation of this talk: V. M. Fomin, V. N. Gladilin, S. N. Klimin, J. Tempere, F. Brosens, E. De Wolf, J. Naudts, D. Callebaut (Universiteit Antwerpen), Y. Bruynseraede, V. V. Moshchalkov, J. Indekeu (KU Leuven), J. H. Wolter (TU Eindhoven).